# Gate Leakage Suppression and Breakdown Voltage Enhancement in p-GaN HEMTs using Metal/Graphene Gates

Guangnan Zhou, Zeyu Wan, Gaiying Yang, Yang Jiang, Robert Sokolovskij, Hongyu Yu, and Guangrui (Maggie) Xia

*Abstract*—In this work, single-layer intrinsic and fluorinated graphene were investigated as gate insertion layers in normally-OFF p-GaN gate HEMTs, which wraps around the bottom of the gate forming Ti/graphene/p-GaN at the bottom and Ti/graphene/SiN$_x$ on the two sides. Compared to the Au/Ti/p-GaN HEMTs without graphene, the insertion of graphene can increase the $I_{ON}/I_{OFF}$ ratios by a factor of 50, increase the $V_{TH}$ by 0.30 V and reduce the off-state gate leakage by 50 times. Additionally, this novel gate structure has better thermal stability. After thermal annealing at 350 °C, gate breakdown voltage holds at 12.1 V, which is first reported for Schottky gate p-GaN HEMTs. This is considered to be a result of the 0.24 eV increase in Schottky barrier height and the better quality of the Ti/graphene/p-GaN and Ti/graphene/SiN$_x$ interfaces. This approach is very effective in improving the $I_{ON}/I_{OFF}$ ratio and gate BV of normally-OFF GaN HEMTs.

## I. INTRODUCTION

Gallium nitride (GaN) possesses excellent physical properties, such as a high critical electric field and a high saturation velocity [1-3], which is ideal for devices with a low specific ON-resistance ($R_{ON}$), a high breakdown voltage and a high operation switching frequency. For switch applications, normally-off transistors are required to provide adequate safety conditions [4-5]. Among many options of normally-off HEMTs [6-8], recessed gate metal-insulator-semiconductor-HEMTs (MIS-HEMTs) and p-GaN gate HEMTs (which lift up the conduction band) have been considered as the two most promising approaches [9-10].

Reducing the gate leakage current remains a big challenge for p-GaN gate HEMTs [11-13]. The forward gate leakage current limits the gate voltage swing and causes drive losses, while the reverse one can lead to the off-state power consumption [14]. Previous studies in MIS-HEMTs revealed that fluorinated graphene can serve as a barrier layer between $Al_2O_3$ and GaN, which contributed to a suppression of gate leakage current by two orders of magnitude [15]. Graphene can act as a strong barrier to atom diffusion and saturate the dangling bonds and defects on the surface [16, 17]. However, there have not been any reports on the effectiveness of graphene in p-GaN HEMTs, which was addressed in this work.

## II. DEVICE STRUCTURE AND FABRICATION

The p-GaN gate HEMTs were fabricated on 100 nm p-GaN/15 nm $Al_{0.2}Ga_{0.8}N$/0.7 nm AlN/4.5 μm GaN epi-structures grown on Si (111) substrates by metal organic chemical vapor deposition (MOCVD) provided by Enkris Semiconductor Inc. Fig. 1 shows the schematic cross-section of the devices. The p-GaN layer was doped with Mg to a concentration of $4 \times 10^{19}$ cm$^{-3}$. The fabrication flow started with p-GaN gate definition by a Cl-based plasma etch followed by a $Cl_2/BCl_3$ plasma etch to form mesas and isolate the devices. The source/drain (S/D) Ohmic contacts were formed by Ti/Al/Ti/Au (20/110/40/50 nm) deposition and annealing was at 830 °C in $N_2$ for 45 s. 120 nm SiN$_x$ deposited by plasma enhanced chemical vapor deposition (PECVD) was used as the first passivation layer. Prior to the deposition of gate metals (40 nm Ti/100 nm Au), single layer graphene grown by chemical vapor deposition (CVD) on Cu foils was transferred to part of the sample surface via the "polymethyl methacrylate (PMMA)-mediated" wet-transfer approach [18]. The undesired part of graphene was etched away by $O_2$ plasma after deposition of gate metal. Finally, the devices were annealed at 350 °C in $N_2$ for 5 minutes to improve the Au/Ti(graphene)/p-GaN interface.

This work was supported by Grant #2019B010128001 and #2017A050506002 from Guangdong Science and Technology Department, Grant #JCYJ20160226192639004 and #JCYJ20170412153356899 from Shenzhen Municipal Council of Science and Innovation. (Corresponding authors: Y. Hong and G. Xia.)

G. Zhou and G. Xia are with and the School of Microelectronics, Southern University of Science and Technology (SUSTech) and the Department of Materials Engineering, the Univerisity of British Columbia (UBC). (e-mail: gxia@mail.ubc.ca)

Z. Wan, Y. Jiang, R. Sokolovskij and H. Yu are with the School of Microelectronics, SUSTech; GaN Device Engineering Technology Research Center of Guangdong, SUSTech; and the Key Laboratory of the Third Generation Semiconductors, SUSTech, 518055 Shenzhen, Guangdong, China. (e-mail: yuhy@sustech.edu.cn)

G. Yang is with the School of Innovation & Entrepreneurship at SUSTech.

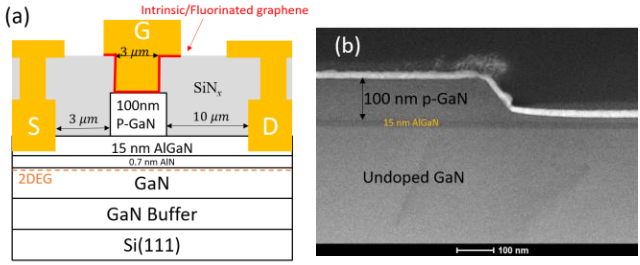

Fig. 1. (a) Schematic of the device structure; (b) Cross-section of the device characterized by a scanning transmission electron microscope (STEM) showing the edge of a p-GaN gate and the p-GaN/AlGaN/GaN structure.

Two types of graphene, intrinsic (I-graphene) and fluorinated graphene (F-graphene), were investigated. F-graphene was realized by exposing the I-graphene to $SF_6$ plasma. In this experiment, the quality and the cleanness of the transferred graphene are crucial, as the metal/semiconductor barrier height heavily depends on the surface states and defects. Especially, the PMMA residuals could lead to Fermi-level pinning at the metal/graphene/p-GaN junction, which would eventually result in the malfunction of these devices [19]. As illustrated in Fig. 2, the Raman spectra and SEM pictures show that the transferred graphene has a high quality ($I_{2D}/I_G \approx 2$) with negligible PMMA contaminations [20].

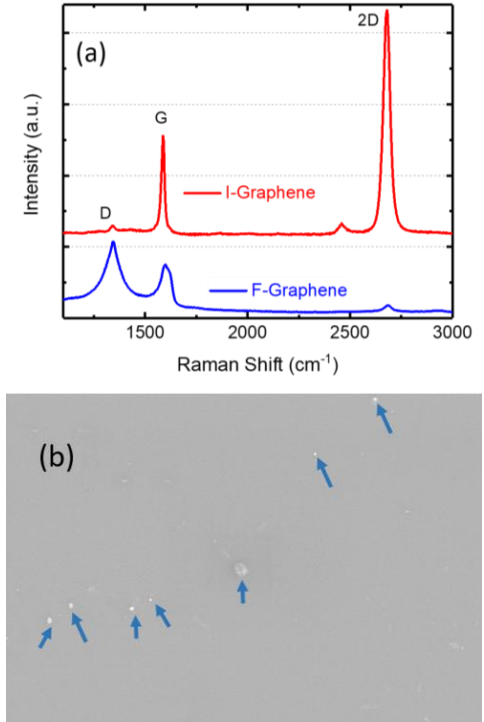

Fig. 2. The transferred graphene characterized by (a) Raman spectroscopy and (b) SEM, a few PMMA residuals are shown intentionally as noted by the arrows.

III. RESULTS AND DISCUSSION

A. HEMTs results

The HEMTs with and without graphene are from the same wafer piece. The gate leakage, transfer and output characteristics of the HEMTs are shown in Fig. 3.

Compared to the HEMTs with Ti/p-GaN gate structures, those with Ti/I-graphene/p-GaN and Ti/F-graphene/p-GaN gate structures have about one order of magnitude lower gate leakage with a forward bias and 2 orders of magnitude with a reverse bias.

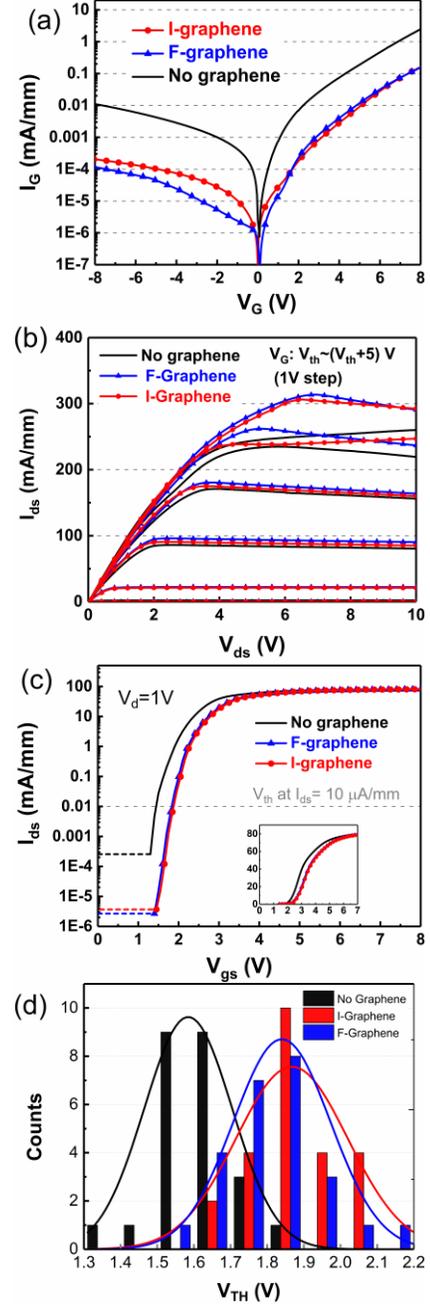

Fig. 3. Comparison of the device performance of the p-GaN HEMTs with I-graphene, F-graphene and without graphene; (a) gate leakage characteristics under $V_{DS}$ = 0 V; (b) output characteristics ($V_G$ from $V_{TH}$ to ($V_{TH}$ + 5V)), 1V step); (c) transfer characteristics under $V_{DS}$ = 1 V, the inset figure is transfer characteristics plotted in linear scale; and (d) $V_{TH}$ uniformity characterization.

Additionally, the HEMTs with graphene have higher threshold voltage ($V_{TH}$) defined at $I_{DS}$ = 10 μA/mm (0.32 V higher for I-graphene and 0.28 V higher for F-graphene). More importantly, the insertion of graphene increased the $I_{ON}/I_{OFF}$ ratio at least one order of magnitude for both types of graphene. Detailed comparison of the extracted DC parameters between three types of devices is shown in Table I. The statistical data of $V_{TH}$ is shown in Fig. 3 (d). The average $V_{TH}$ values of the devices without graphene, with I-graphene and with F-graphene are 1.58 V, 1.86 V and 1.84 V, respectively. The increase of $V_{TH}$ should be attributed to higher Schottky barrier height ($\Phi_B$) and p-type doping in graphene introduced in the "PMMA-mediated" wet-transfer process [18]. In summary, the insertion of graphene reduced the gate leakage current and increased $V_{TH}$ without sacrificing any output performances.

TABLE I
COMPARISON OF EXTRACTED DC PARAMETERS

| Parameters | No graphene | I-Graphene | F-Graphene |
|---|---|---|---|
| $I_G$ ($V_G$ = 8V) (mA/mm) | ~ 2 | ~ 0.1 | ~ 0.1 |
| $I_{OFF}$ (mA/mm) | $2.5 \times 10^{-4}$ | $3.6 \times 10^{-6}$ | $1.2 \times 10^{-6}$ |
| $V_{TH}$ (V) | 1.54 | 1.86 | 1.82 |
| $R_{ON}$ (Ω•mm) | 12.4 | 12.8 | 11.9 |
| Gate BV (V) | 9.8 | 12.05 | 12.0 |

*B. Effects of Graphene during Annealing Treatment*

In practice, passivation layers are necessary for metal pads and interconnect protection. This means that the gate metal/p-GaN interfaces will commonly experience some low-temperature thermal annealing. For the gate-metal-first process, the interface will undergo an even higher thermal budget for S/D contact annealing [13]. To investigate the effects of graphene during these thermal steps, the gate leakage of devices had also been measured before the annealing treatment (350 °C in $N_2$ for 5 minutes). Fig. 4 presents the gate leakage characteristics and gate breakdown voltage (BV) before and after annealing.

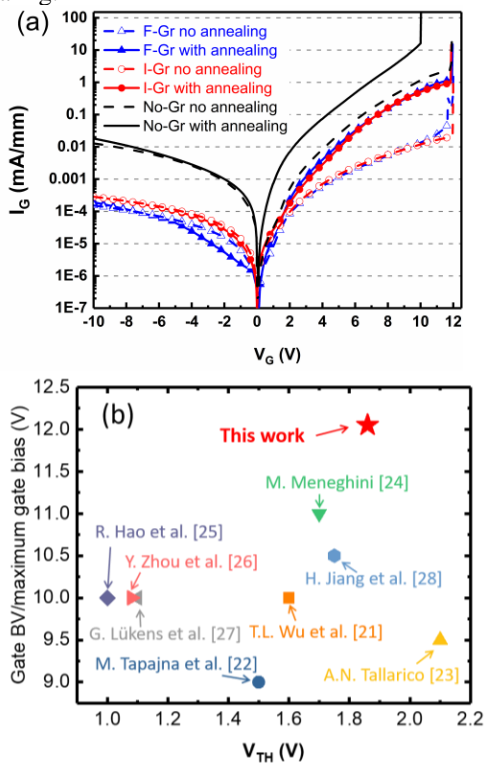

Fig. 4. (a) Comparison of the gate leakage current and gate breakdown voltage before and after 350 °C 5 min annealing; (b) Gate BV or maximum gate bias versus threshold voltage for p-GaN gate HEMTs fabricated by several groups.

As illustrated in Fig. 4, in the case of forward $V_G$ bias, all the devices have a considerable increase in gate leakage. This should be due to the reduction of $\Phi_B$. After the annealing, the gate BV of the devices without graphene was 9.80 V whereas the devices with graphene broke down at 12.05 V. To our knowledge, this is the highest gate BV that has been reported in HEMTs with p-GaN Schottky gates. Fig. 7 summarize the gate BV or maximum gate bias versus $V_{TH}$ for p-GaN gate HEMTs fabricated by several groups. Compared with other reported p-GaN gate HEMTs, our devices with graphene insertion exhibit the highest gate BV and a relative high $V_{TH}$. These improvements are crucial for p-GaN gate HEMT device for power switching applications.

To get further insights into the role of graphene on Ti/p-GaN interfaces, scanning transmission electron microscopy (STEM) images of the HEMTs were obtained. As shown in Fig. 5, the Ti/p-GaN interface quality has been much improved by the insertion of graphene. The thin darker regions pointed out by the yellow arrows in Fig. 5(a) were confirmed to be small voids or TiN islands by energy dispersive spectroscopy (EDS), as illustrated in the insets of Fig. 5(a). The inset figure on the left shows that the darker region has TiN, while the right one illustrates a void where only background signals/noise exist in EDS. Meanwhile, the interface with graphene is more uniform and cleaner. This means that the graphene interlayer helped to reduce the formation of TiN and interface voids and maintain a more stable Ti/graphene/p-GaN interface during annealing, contributing to a lower leakage current and a higher gate BV.

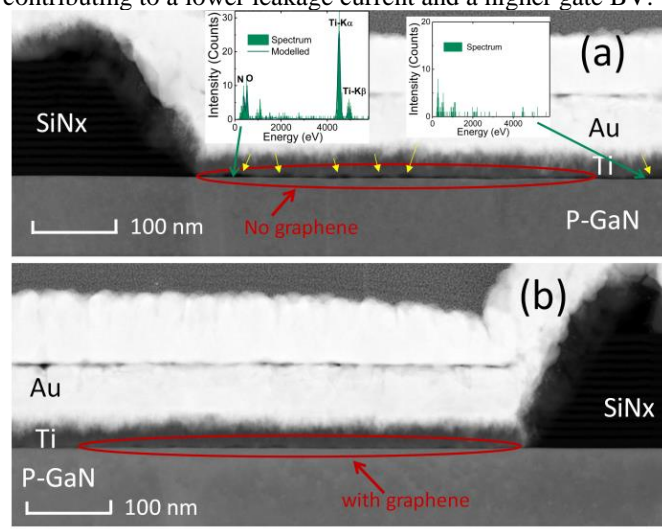

Fig. 5. Cross-section STEM of the HEMTs after 350 ºC 5 min annealing in the gate region (a) without graphene; Insets: EDS show the darker regions are either TiN or voids; (b) with I-graphene. The Au/Ti/p-GaN interface has been much improved by graphene.

*C. Mechanism Study with Ti/graphene/p-GaN Schottky diodes*

Besides the material analysis, Au/Ti/graphene/p-GaN Schottky contacts have been fabricated as seen in the inset of Fig. 6(a). Considering the fact that the I-graphene and F-graphene had similar effects in the previous experiments, only I-graphene was studied here. This test structure consists of two Schottky contacts back-to-back [29]. The I–V measurements were carried out in the bias range of 0 to 6 V. Before annealing, the systems with and without I-graphene have comparable current density, whereas they are significantly different after 350 °C 5 minutes annealing.

The temperature dependence of the I-V characteristics has also been studied for these systems to extract $\Phi_B$. Taking the Mg doping concentration of $4 \times 10^{19}$ cm$^{-3}$ into consideration, the value of $E_{00}/kT$ is in the range between 0.85 and 1.34 in the examined temperature range. This implies that thermionic field emission (TFE) should be the dominant mechanism in our systems. Hence, the expression of the TFE reverse current was used to fit our experimental data [13, 30]

$$I_s = \frac{A_c A^* T \sqrt{\pi E_{00}}}{k} \cdot \sqrt{q(V - V_n) + \frac{q\Phi_B}{\cosh^2\left(\frac{E_{00}}{kT}\right)}}$$
$$\cdot \exp\left(-\frac{q\Phi_B}{E_{00} \coth\left(\frac{E_{00}}{kT}\right)}\right) \quad (1)$$

$$I_{TFE} = I_s \cdot \exp\left(\frac{qV}{kT} - \frac{qV}{E_{00} \coth\left(\frac{E_{00}}{kT}\right)}\right) \quad (2)$$

$$E_{00} = \frac{qh}{4\pi}\sqrt{\frac{N_A}{m^* \varepsilon}} \quad (3)$$

where $V$ is the applied voltage, $A_c$ is the contact area, $A^*$ is the Richardson constant, $q$ is the elementary charge, $k$ is the Boltzmann constant, $h$ is Planck's constant, $T$ is the absolute temperature, and $\varepsilon$ and $m^*$ are the dielectric constant for GaN and the effective mass for holes, respectively. $\Phi_B$ represents the Schottky barrier height of the metal(graphene)/p-GaN interface. In our calculation, we used the values of $\varepsilon = 8.9\,\varepsilon_0$, $A^* = 27.9$ A/cm$^2$K$^2$, and $m^* = 0.81\,m_0$.

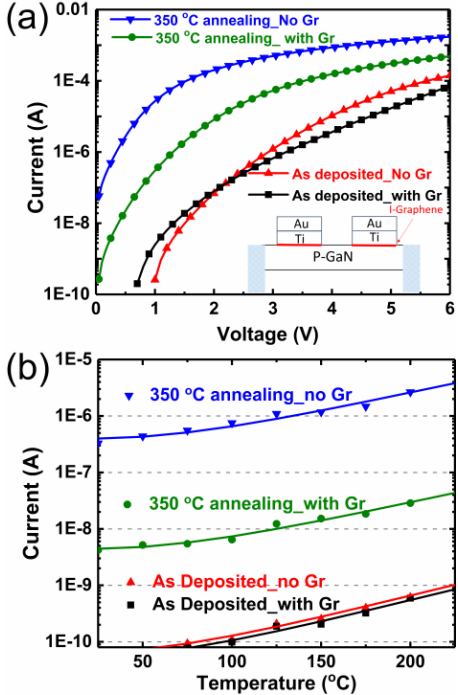

Fig. 6. (a) I–V characteristics acquired in back-to-back diodes. Inset: Schematics of the sample structure; (b) Temperature dependence of the current measured at V = +0.2 V in back-to-back Schottky contacts. The symbols are measured data. The lines are the fitting curves.

The extracted $\Phi_B$ from the data (Fig. 6(b)) is 2.08 eV for the contacts without graphene and 2.09 eV for those with I-graphene, respectively. After annealing the sample at 350 °C in N$_2$ for 5 min, $\Phi_B$ decreased to 1.65 eV (no Gr) and 1.89 eV (with graphene), which results in approximately one order of magnitude smaller current in the sample with graphene. The extracted $\Phi_B$ are summarized in Table II. These data are consistent with the experimental value reported in the literature for Ti/p-GaN interfaces [13].

It is worth mentioning that this back-to-back Schottky diodes experiment was conducted on p-GaN wafer without any passivation or etching process. The metal/p-GaN interfaces in a HEMT can be more defective due to the ion bombardment during the fabrication process (e.g. passivation deposition and gate dielectric etching).

### D. Analysis and Modeling of Gate Leakage Current

In this part, the gate current characteristics in Fig. 4(a) and the corresponding mechanism will be analyzed more thoroughly. The data will be fitted to a physical model from Ref. [31]. The trap depth and the corresponding prefactors of the HEMTs with/without graphene will be extracted.

As shown in Fig. 7, the gate structure includes two junctions, i.e. the metal (graphene)/p-GaN junction (Schottky junction) and the p-GaN/AlGaN/GaN junction (p-i-n junction). The gate leakage mechanisms are different in different gate bias regimes.

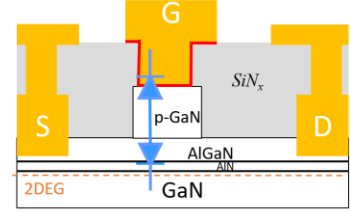

Fig. 7. Schematic representation of the Au/Ti/p-GaN/AlGaN/GaN HEMT and its equivalent circuit.

In the V$_G$ < 0 regime, vertical gate current is blocked due to the reverse bias of the p-i-n junction. An electron current is likely from the Schottky metal near the gate edges through a leakage path along the p-GaN sidewall and these electrons are emitted into the 2DEG layer. As shown in Fig. 4(a), the I$_G$ with/without graphene at reverse bias had negligible change after annealing even though the Schottky barrier changed. This further confirms that the dominant factor for I$_G$ at the negative bias is associated with the surface defects caused during the fabrication process. A similar conclusion has also been reported for previous p-GaN gate HEMTs [31, 32]. According to Ref [31, 32], the dominant gate leakage mechanism associated with p-GaN surface defects is found to be Poole–Frenkle emission (PFE) for the case of V$_G$ < 0 as shown in Fig. 7(b), which can be described by:

$$J = C\,E \exp\left(-\frac{q(\Phi_t - \beta\sqrt{E})}{kT}\right), \quad (4)$$

where $C = qn\mu_n$, $\beta = \sqrt{q/\pi\epsilon}$ is the Schottky factor, $k$ is Boltzmann's constant, $q\Phi_t$ is the trap depth, and $E = V_{GS}/L_{GS}$ is the electric field, with $L_{GS}$ the shortest distance between the gate and source metal.

The temperature dependent I$_G$-V$_G$ of the HEMTs after annealing have been measured from 300 K to 450 K. The PFE model has been used to fit the data in the negative bias, with the trap depth $q\Phi_t$ and prefactor C as the fitting parameters (Fig. 8). The fitting results are illustrated in Fig. 8, where the fitting curves are plotted as solid lines and symbols represent selected measurement points. The trap depths are $q\Phi_{t,I-Gr} = 0.199\ eV$ and $q\Phi_{t,no-Gr} = 0.165\ eV$, respectively. The prefactor C in the HEMTs without graphene is $\frac{C_{no-Gr}}{C_{I-Gr}} = \frac{585.70}{47.46} = 12.34$ times of that with I-graphene. This explains why $I_G$ in the HEMTs with/without graphene differ a lot in the as-deposited case even though their $\Phi_B$ are comparable extracted in Section II-C. Graphene layers passivated the p-GaN surface resulting in the neutralization of interface states and, consequently, less electron leakage toward the gate foot along the p-GaN surface.

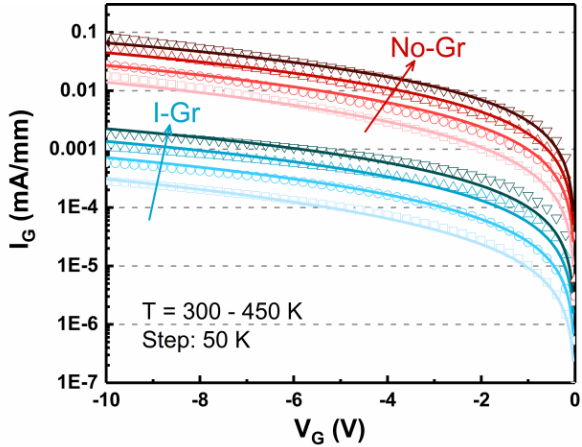

Fig. 8. Negative gate bias current from T = 300 to 450 K measured on HEMTs with I-graphene and without graphene. The symbols are selected measured data. The lines are the fitting curves using the PFE model.

TABLE II
SUMMARY OF MODELLING RESULTS OF THE BACK-TO-BACK DIODES AND THE HEMTS GATE CURRENT

|  |  | As Deposited | 350 ºC 5 min annealing |
|---|---|---|---|
| $\Phi_B$ (eV) in diodes | With I-graphene | 2.08 | 1.89 |
|  | No graphene | 2.09 | 1.65 |
| $\Phi_t$ (eV) in HEMTs | With I-graphene | \ | 0.199 |
|  | No graphene | \ | 0.165 |
| $C$ ($\Omega^{-1}$) in HEMTs | With I-graphene | \ | $5.86 \times 10^{-7}$ |
|  | No graphene | \ | $4.75 \times 10^{-8}$ |

In the $V_G > 0$ regime, the p-i-n junction will be positive bias while the Schottky junction will be reverse bias. In the case of no-annealing, it is found that the $I_G$ are comparable in the regime of positive and negative gate bias in Fig 4(a). We deduce that the vertical gate current is also small since the Schottky barrier is large enough. The graphene contributes to a lower gate leakage by reducing the interface defects similar to the $V_G < 0$ regime. In the case of after-annealing, the gate leakage current in positive bias is much larger than in negative bias. The dominant current shall be from the vertical gate current due to a lower $\Phi_B$ in the Schottky junction. The voltage dropped at the p-i-n junction is large enough to turn on the diode and the dominant conduction mechanism is found to be thermionic field emission. Thus, graphene helps to reduce the gate leakage current by maintaining a more stable interface and a higher $\Phi_B$ after annealing as analyzed in Section II-C.

Based on these evidences, it can be deduced that graphene can help reduce the interface defects and maintain a more stable Au/Ti/p-GaN interface during annealing. In $V_G > 0$ regime, the larger $\Phi_B$ contributes to a lower gate leakage current and the higher $V_{TH}$. Graphene can also prevent the decrease of gate BV during annealing. In the $V_G < 0$ regime, as the defects at the p-GaN surfaces during the fabrication process are the dominant factor for gate leakage, graphene reduces the gate leakage by saturating the defects at the p-GaN surface. According to fitting the experimental data, it is found that graphene changes the trap depth $q\Phi_t$ from 0.165 eV to 0.199 eV and reduces the prefactor C by 12.3 times.

IV. CONCLUSION

In this work, intrinsic and fluorinated graphene were investigated as gate insertion layers for normally-OFF p-GaN HEMTs, which formed Au/Ti/graphene/p-GaN interfaces in the middle and Au/Ti/graphene/ $SiN_x$ on the two sides. 50 times larger $I_{ON}/I_{OFF}$ ratios, 0.30 V higher $V_{TH}$ increase, 50 times off-state gate leakage reduction have been achieved by the insertion of graphene. For the first time, 12.1 V gate BV has been achieved with I-graphene for Schottky gate p-GaN HEMTs. This is considered to be the result of a 0.24 eV higher $\Phi_B$, and the better graphene/p-GaN interfaces. In the negative gate bias, $I_G$ has been fitted to PFE model, which revealed that graphene layers contribute to a 0.034 eV larger trap depth and a 12.3-time smaller prefactor. As single-layer graphene can be prepared in wafer-size areas [33], This approach is mass-production compatible and very effective in improving the $I_{ON}/I_{OFF}$ ratios and increasing $V_{TH}$ and gate BV of p-GaN gate HEMTs.

ACKNOWLEDGMENT

The work was conducted at Materials Characterization and Preparation Center (MCPC) at SUSTech, and we acknowledge the technical support from the staff and engineers at MCPC. Dr. Mengyuan Hua from SUSTech is acknowledged for useful discussions.